\documentclass[slac_one]{revtex4}
\usepackage{graphicx}
\usepackage{fancyhdr}
\begin{document}

\title{Light Hadron Physics at the B Factories} 

\author{Selina Z. Li (from the BaBar Collaboration)}
\affiliation{Stanford Linear Accelerator Center, Menlo Park, CA 94025, USA}

\begin{abstract}
We report measurements of hadronic final states produced in $e^+e^-$ annihilations from the BaBar and Belle experiments.  In particular, we present cross sections measured in several different processes, including two-photon physics, Initial-State Radiation, and exclusive hadron productions at center-of-mass energies near $10.58$ GeV.  Results are compared with theoretical predictions.
\end{abstract}

\maketitle

\thispagestyle{fancy}
\section{INTRODUCTION} 
The BaBar and Belle experiments are designed mainly to study $CP$-violation in B decays produced in $e^+e^-$ collisions.  However, as a by-product, they can also be used to study other hadronic final states, some of which will be covered in this proceeding.  We will present some of the experimental results obtained from analyses of two-photon physics, Initial-State Radiation (ISR), and exclusive hadron production at center-of-mass (CM) energies at or near 10.58 GeV.  Results presented here are based on $95-427$ fb$^{-1}$ data samples collected by the BaBar and the Belle detectors, whose detailed descriptions can be found in Ref.~\cite{ref:BaBar} and \cite{ref:Belle}, respectively. 
\section{TWO-PHOTON PHYSICS: THE $\gamma\gamma\to\pi^0\pi^0$ REACTION}
\begin{figure}[htbp]
\centering
\begin{tabular}{ccc}
\includegraphics[width=50mm]{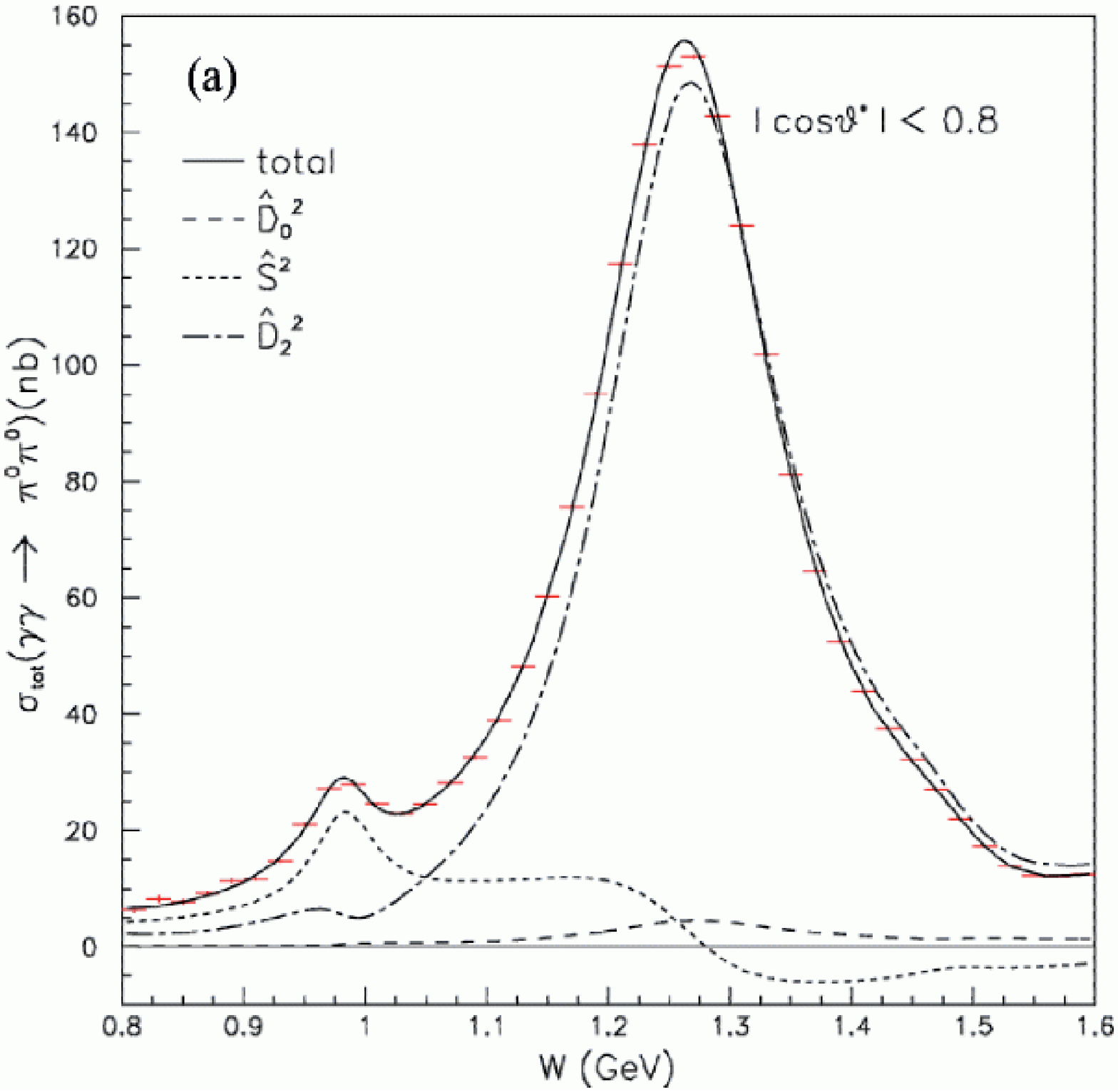} & \includegraphics[width=50mm]{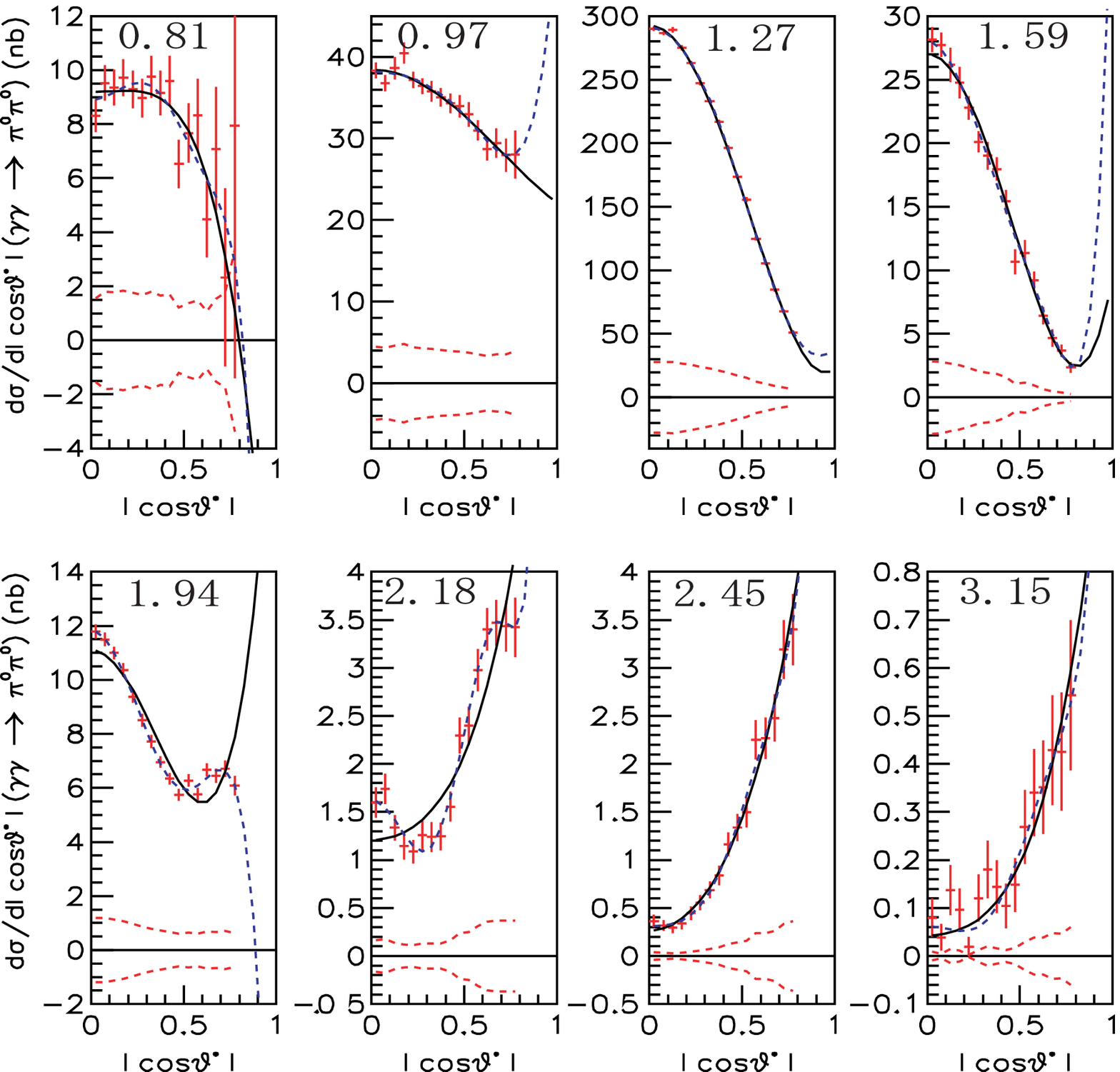} & \includegraphics[width=60mm]{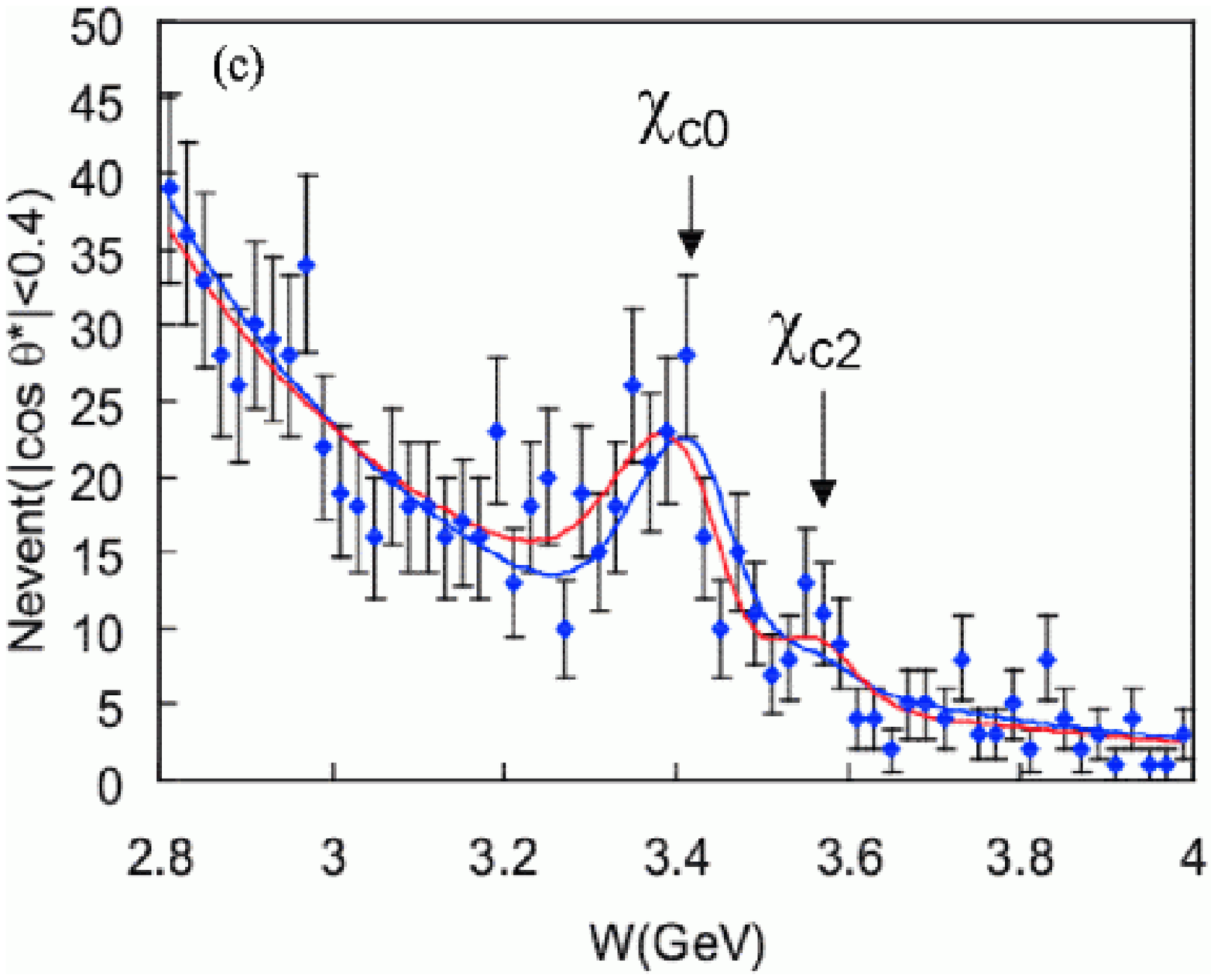}
\end{tabular}
\caption{Belle's $\gamma\gamma\to\pi^0\pi^0$: (a) total cross section; (middle) differential cross section with SD-wave (solid) and SDG-wave (dashed) fits; (c) fits including $\chi_{c0}$ with and without interference for $\chi_{c2}$.}
\label{fig:Belle_pi0pi0}
\end{figure}

Two-photon reactions resulting from the process $e^+e^-\to e^+ e^- \gamma \gamma$  can be used to study resonance structure and to test QCD models.  Belle analyzed a $95$ fb$^{-1}$ data sample to study the process $\gamma\gamma\to\pi^0\pi^0$ in a no-tag analysis~\cite{ref:twophoton}, where the $e^+e^-$ scatter at very small angles and escape undetected.  In the region $0.8-1.6$ GeV, the production cross sections (Fig.~\ref{fig:Belle_pi0pi0}a) as a function of the CM energy ($W$) of the $\pi^0\pi^0$ system are fitted to obtain information on the S and D waves where two resonant structures $f_0(980)$ and $f_2(1270)$ contribute. The angular dependence of the differential cross sections shows a large-scattering-angle enhancement at low energies ($W<1.7$ GeV) and a forward angle peak at high energies ($W>2.4$ GeV).  The fits suggest that a G-wave becomes necessary at energies above $1.9$ GeV.  
Belle recently has analyzed a $223$ fb$^{-1}$ data sample to study the higher energy region~\cite{ref:pi0pi0} where a $\chi_{c0}$ charmonium is observed with a significance of more than seven standard deviations (Fig.~\ref{fig:Belle_pi0pi0}c).  The ratio of cross sections for neutral to charged pion-pair production~\cite{ref:pipi} is almost constant at $W>3.1$ GeV and is significantly larger than the prediction of leading order QCD.  This ratio is compared with QCD models for hadron pair production listed in Table~\ref{Table:pi0pi0}.
\begin{table}[htbp]
\begin{center}
\caption{Comparison of the ratio of cross sections for neutral to charged pion-pair production}
\begin{tabular}{|l|c|c|c|c|}
\hline & \textbf{Leading Term QCD~\cite{ref:leading}} & \textbf{pQCD~\cite{ref:pQCD}} & \textbf{``Handbag'' Model~\cite{ref:handbag}} &\textbf{Belle preliminary results}
\\
\hline $\sigma(\pi^0\pi^0)/\sigma(\pi^+\pi^-)$& $0.03-0.07$ & $0.1$ & $0.5$ & $0.32\pm0.03\pm0.05$\\
\hline
\end{tabular}
\label{Table:pi0pi0}
\end{center}
\end{table}
\section{STUDIES OF INITIAL-STATE RADIATION PHYSICS AT $\Upsilon$(4S) ENERGIES}
In an ISR event, either the electron or positron emits an ISR photon, lowering the effective invariant mass of the hadronic system.  This allows us to measure the hadronic cross section of $e^+e^-\to$ hadrons from threshold to CM energy of $\sim5$ GeV.  Therefore, ISR physics gives access to hadron production over a continuous, wide energy range in a single experiment.  The high integrated luminosity allows BaBar to be competitive with $e^+e^-$ experiments at lower CM energy region such as CMD-2 and SND, and to extend the measurements to higher masses.  Experimentally, we require the event to have an energetic ISR photon with energy greater than 3 GeV, and then fully reconstruct the hadronic final states.  Selecting ISR photon within the fiducial region of the detector also forces the recoil hadronic system to be within the fiducial region.  In the next subsections, we will discuss measurements done for ISR physics based on $232$ fb$^{-1}$ of BaBar data.  
\subsection{ISR: The $K^+K^-\pi^0$, $K_s^0K^{\pm}\pi^{\mp}$, $K^+K^-\eta$  Final States}
\begin{figure}[htbp]
\centering
\begin{tabular}{ccc}
\includegraphics[width=40mm]{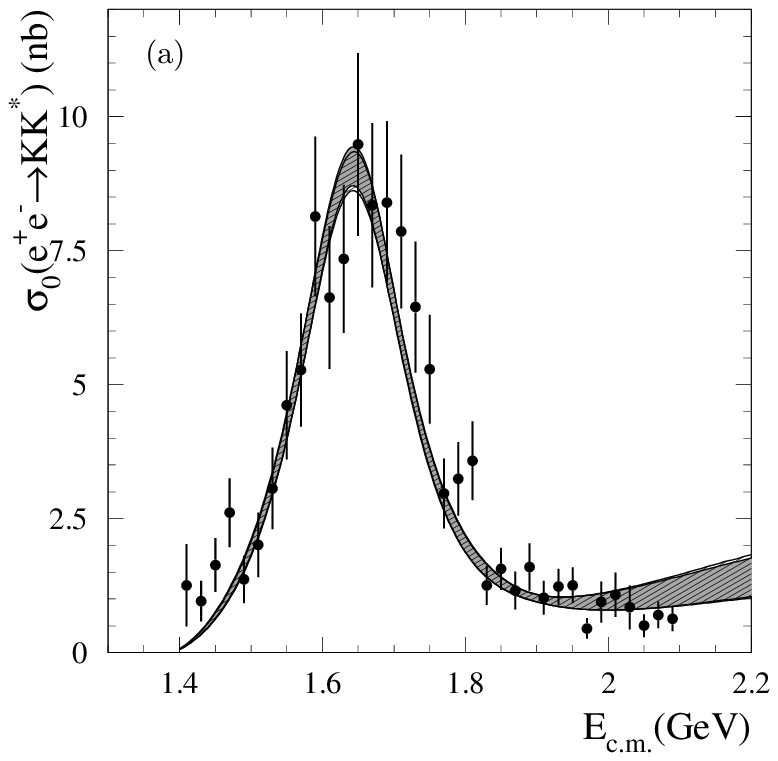} & \includegraphics[width=40mm]{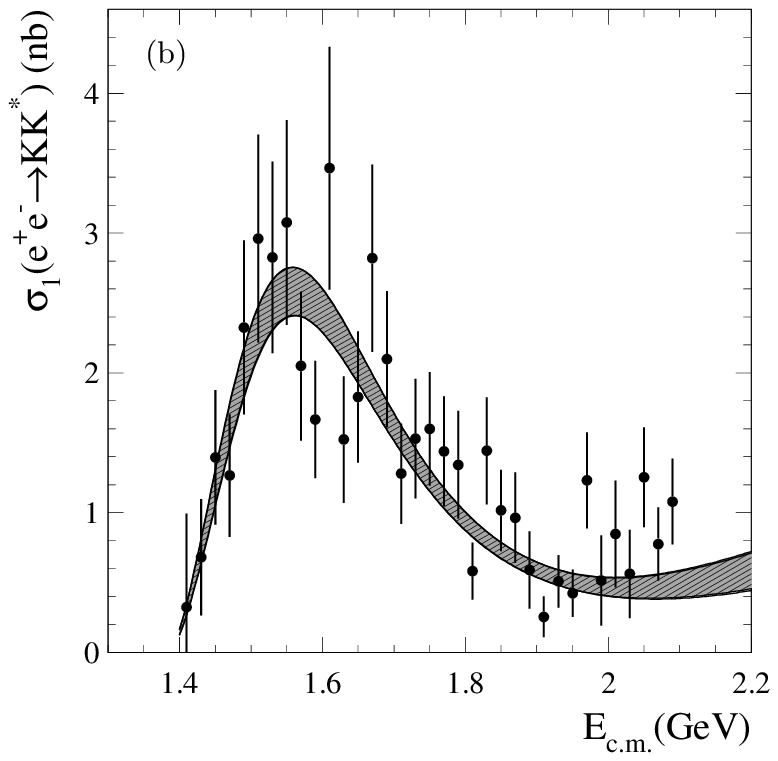} &\includegraphics[width=45mm]{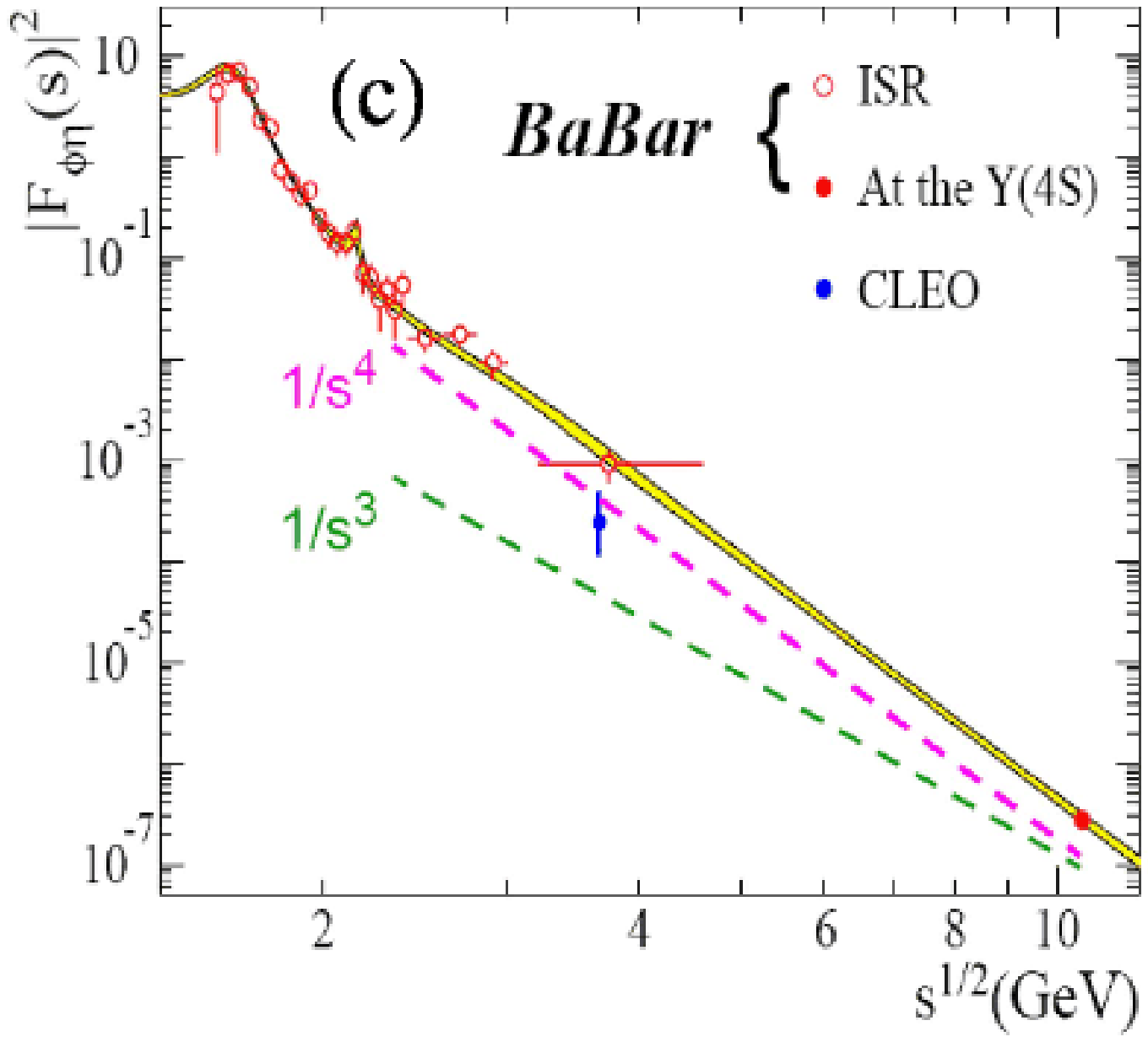} 
\end{tabular}
\caption{The cross sections for (a) isoscalar and (b) isovector components of $KK\pi$ from BaBar data; (c) transition form factors as a function of energy combined with other measurements to compare with theoretical predictions and the solid band is the result of a fit~\cite{ref:fit}.}
\label{fig:BaBar_KKpi}
\end{figure}

From a Dalitz plot analysis of the $KK\pi$ final state~\cite{ref:KKpi}, we can separate the isoscalar and isovector components and measure their cross sections (Fig.~\ref{fig:BaBar_KKpi}a,b).  The isoscalar and isovector channels are dominated by resonances that are consistent with $\phi(1680)$ and $\rho(1450)$.
In the $K^+K^-\eta$ final state~\cite{ref:KKpi}, the main contribution is from $\phi\eta$ whose cross section is converted to transition form factors as a function of energy (Fig.~\ref{fig:BaBar_KKpi}c).  The result from this ISR measurement combined with that of the exclusive production of $e^+e^-\to\phi\eta$ at $10.58$ GeV~\cite{ref:phieta} shows a strong preference for an energy dependence $1/s^4$ of the cross section as predicted in Ref.~\cite{ref:s4} rather than the prediction of $1/s^3$ from Ref.~\cite{ref:s3}.  
\subsection{ISR: The $\pi^+\pi^-\pi^0\pi^0$ Final State}
The channels with four pions in the final state dominate the hadron cross section in the $1-2$ GeV range where a precision measurement is needed for an improved determination of the hadronic contribution to the anomalous magnetic moment of the muon $a_\mu=(g-2)_\mu/2$ and of the running fine structure constant at the $Z^0$-pole $\alpha(m_Z^2)$.  The reduction of uncertainty in this mode is therefore an important contribution to the overall precision of $a_\mu$~\cite{ref:g2}.  The cross section is consistent with previous results, especially with SND at low energy region ({Fig.~\ref{fig:Babar_4pi}b).  
\begin{figure}[htbp]
\centering
\begin{tabular}{ccc}
\includegraphics[width=35mm]{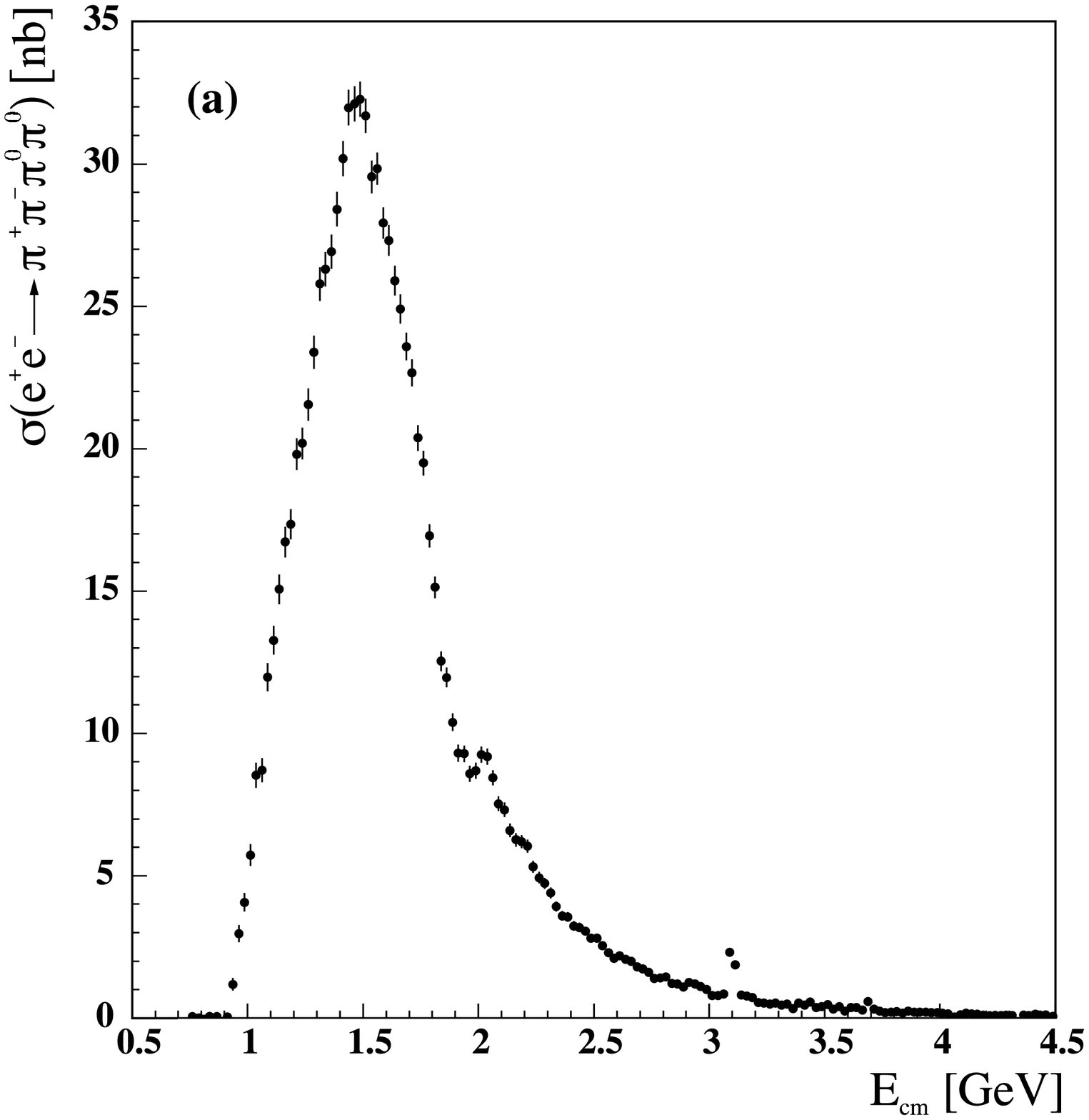} & \includegraphics[width=35mm]{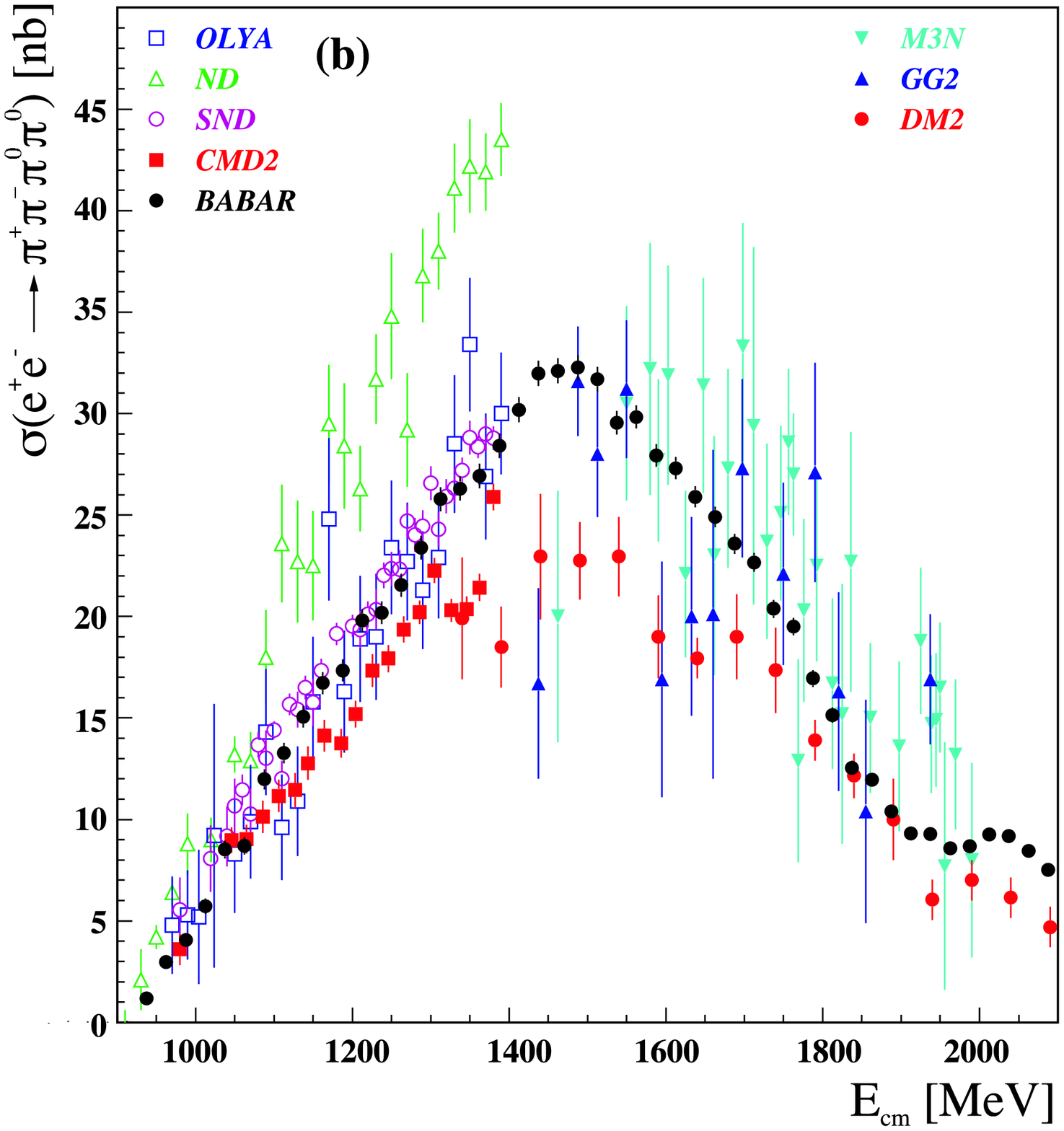} & \includegraphics[width=40mm]{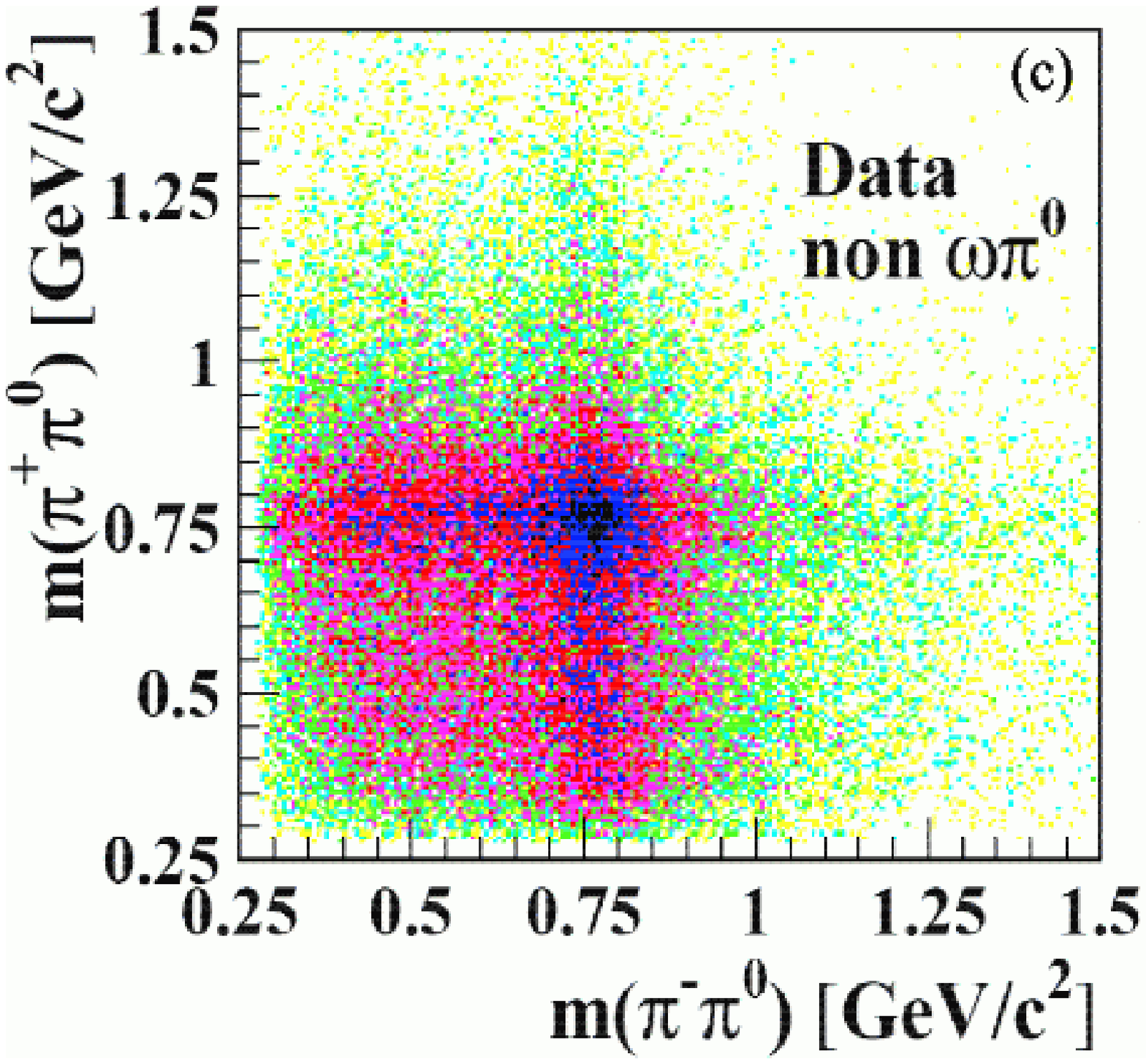}
\end{tabular}
\caption{BaBar's $\pi^+\pi^-\pi^0\pi^0$: (a) cross section; (b) zoom-in at low energy to compare with other experiments; (c) scattering plot shows the $\rho^+\rho^-$ peak.}
\label{fig:Babar_4pi}
\end{figure}
This is the first measurement above $2.5$ GeV.  The resonant structure observed here is dominated by the $\omega\pi^0$ and $a_1(1260)\pi$ channels, along with evidence of $\rho^+\rho^-$ ({Fig.~\ref{fig:Babar_4pi}c).  The latter channel is also seen in exclusive production of $e^+e^-\to\rho^+\rho^-$ at $10.58$ GeV as discussed later in Section~\ref{sec:excl_rhorho}.  The preliminary precision of $8\%$ (estimated to eventually be $\sim5\%$ over the peak region) will help improve the precision uncertainty of the hadronic contribution to $a_\mu$.  
\section{EXCLUSIVE HADRON PRODUCTION AT 10.58 GEV} \label{sec:excl}
Exclusive hadron production from $e^+e^-$ at $10.58$ GeV without ISR attracts little attention due to the expectation of low rates, but there can be surprises.  With the high integrated luminosity collected at BaBar, the first observations of the final states $e^+e^-\to\rho^0\rho^0$ and $\phi\rho^0$~\cite{ref:phirho} with $C=-1$, have established a previous unobserved new source of rare hadronic events through the Two-Virtual-Photon-Annihilation (TVPA) process. This opened a new avenue for the study of hadron production mechanism.  Many other rare exclusive modes have been observed since then, some from TVPA process while others through single photon annihilation.  In the next two subsections, we will concentrate on BaBar's new results of final states allowed via single-virtual-photon annihilation.  
\subsection{The $e^+e^-\to p\bar{p}p\bar{p}$ Reaction} \label{sec:excl_pppp}
The strategy in studying the exclusive hadron production such as $e^+e^-\to p\bar{p}p\bar{p}$ is to fully reconstruct the hadronic final states and then to require the invariant mass of the hadronic system to have the nominal CM energy within resolution requirements.  BaBar has made the first observation of a four-proton exclusive final state in $e^+e^-$ annihilations from $427$ fb$^{-1}$ of data (Fig.~\ref{fig:Babar_4p}).  In the scatter plot Fig.~\ref{fig:Babar_4p}b, the $p\bar{p}$ mass spectrum is peaking near threshold.  Baryon production in PYTHIA (JETSET) shown in Fig.~\ref{fig:Babar_4p}c actually models this quite well which appears to be another success of the string fragmentation model.  We performed a full angular analysis on the signal events and compare the results to several different hadronization models.  For these models we have, JETSET is the closest match to the data, but there are still other features in the data pointing to imperfections of the hadronization models.
\begin{figure}[htbp]
\centering
\begin{tabular}{ccc}
\includegraphics[width=45mm]{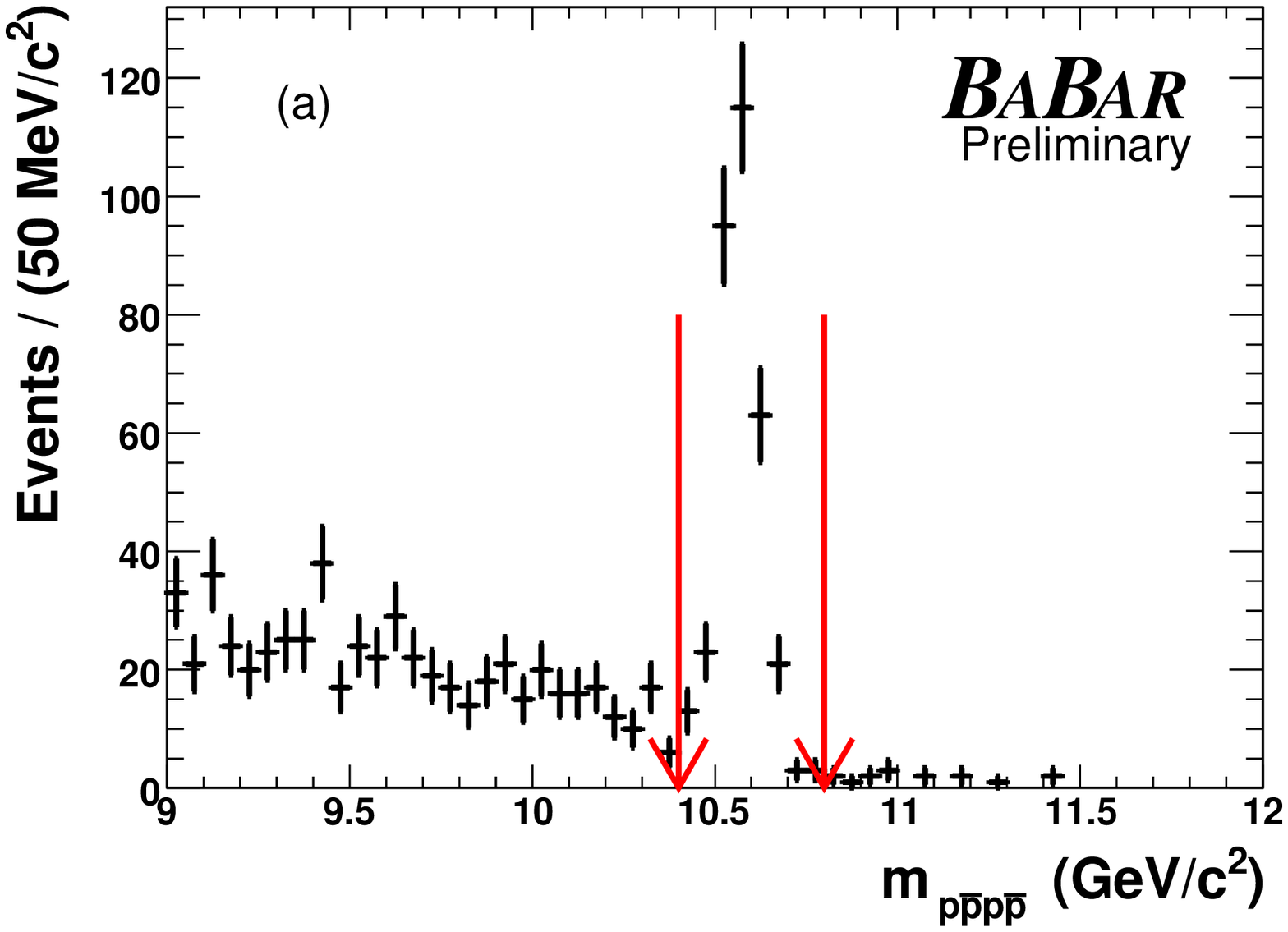} & \includegraphics[width=45mm]{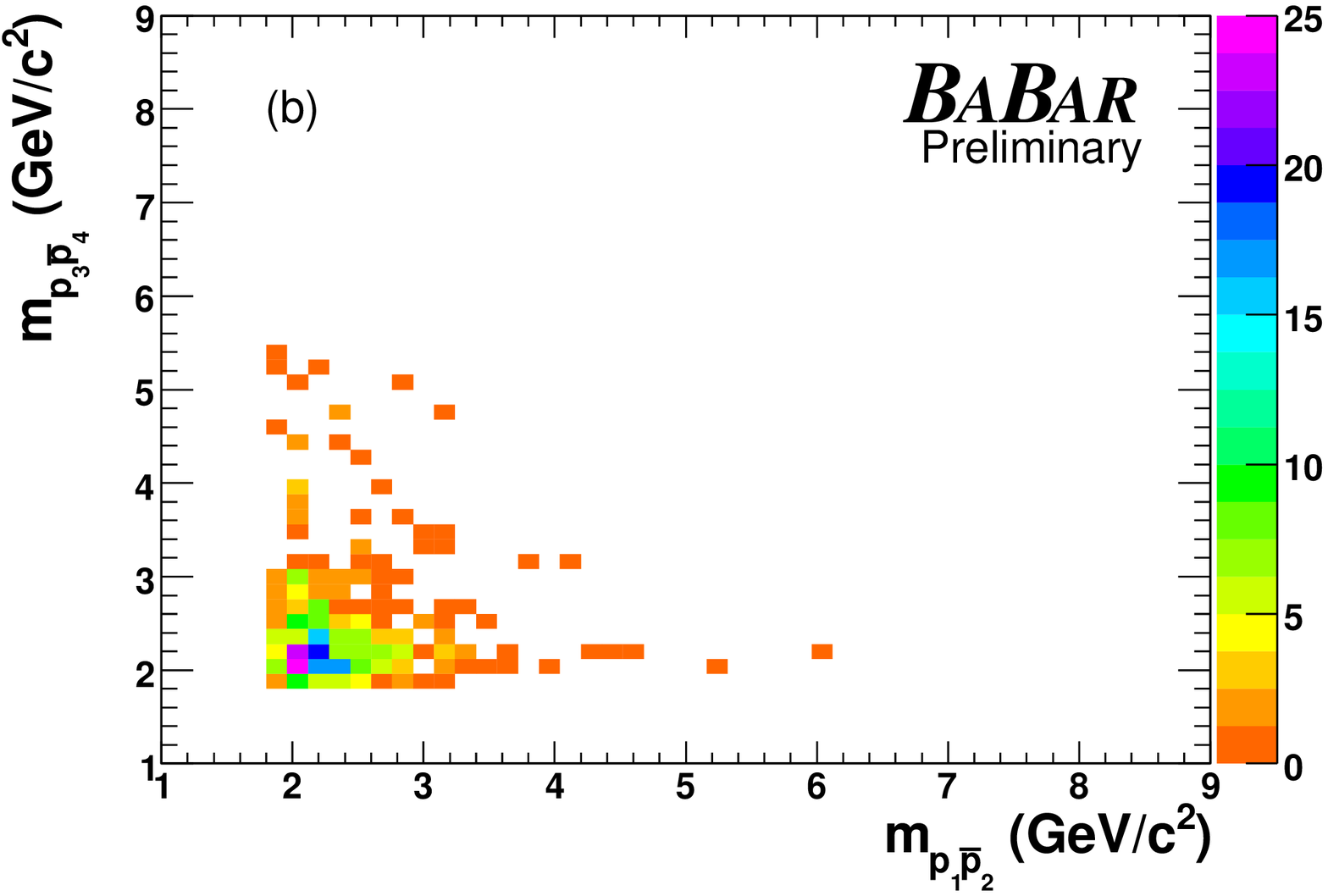} & \includegraphics[width=45mm]{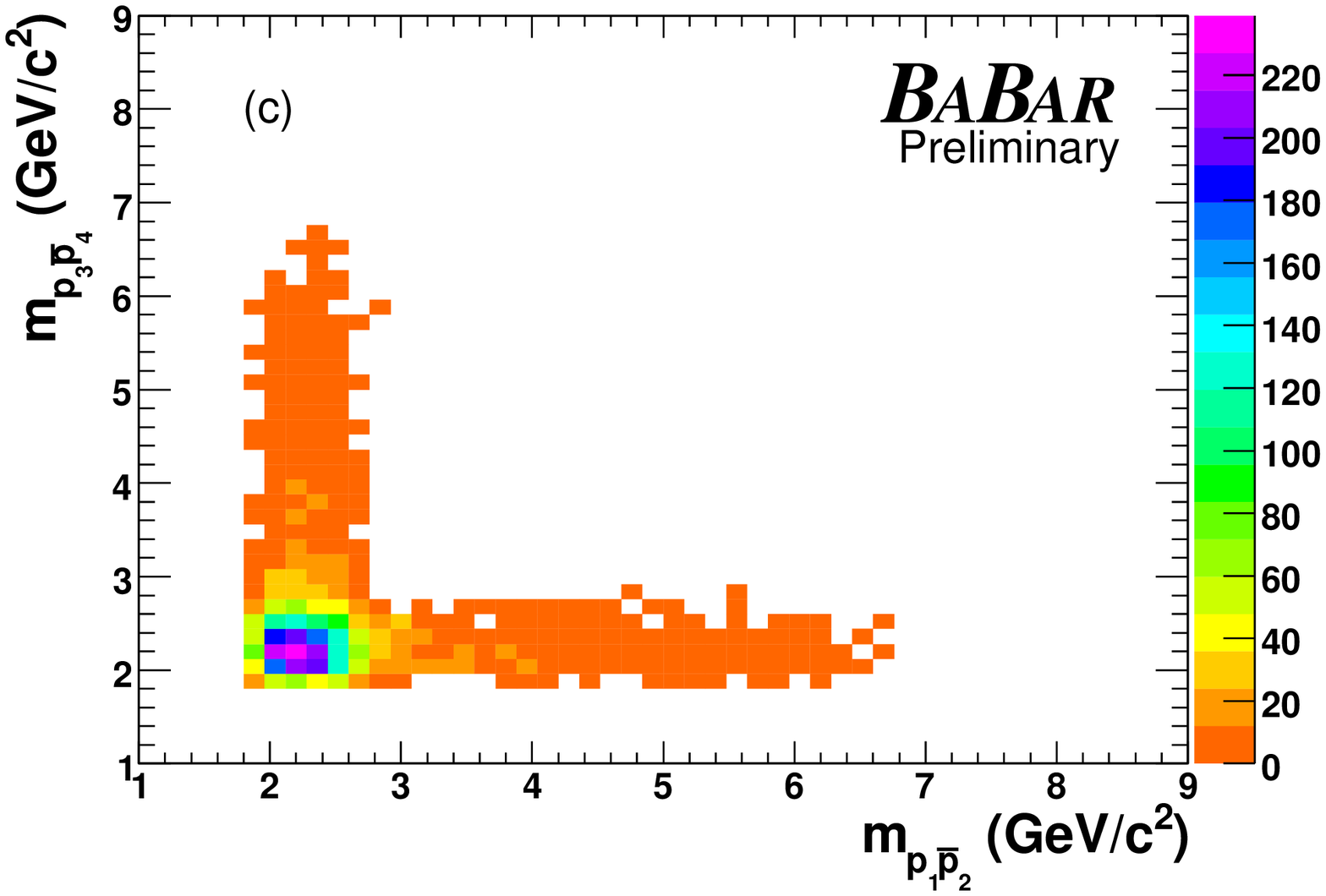}
\end{tabular}
\caption{BaBar's $p\bar{p}p\bar{p}$: (a) invariant mass of $p\bar{p}p\bar{p}$ with vertical lines indicating the mass requirement; (b) data; (c) JETSET signal Monte Carlo sample.}
\label{fig:Babar_4p}
\end{figure}
\subsection{The $e^+e^-\to\rho^+\rho^-$ Reaction} \label{sec:excl_rhorho}
BaBar has also made the first observation of $e^+e^-\to\rho^+\rho^-$~\cite{ref:rhoprhom} based on $379$ fb$^{-1}$ of data from which events with exactly two well reconstructed, oppositely charged tracks and two $\pi^0$ candidates are selected.  We require the invariant mass of the four pions to be within the CM beam energy.  The mass scatter plot shows strong peaks at the $\rho$ mass.  A 2D fit yields $329\pm25$ $e^+e^-\to\rho^+\rho^-$ events with a total cross section measured to be $19.5\pm1.6(stat.)\pm3.2(sys.)$ fb.  The large, clean sample of signal events allows us to perform an angular analysis to test QCD at the amplitude level.  Assuming a one-photon production mechanism, this vector-vector final state can be described by three independent helicity amplitudes, $F_{00}$, $F_{10}$, and $F_{11}$.  Fits to the angular distributions with the normalization constraint $|F_{00}|^2+4|F_{10}|^2+2|F_{11}|^2=1$ reveal that $|F_{00}|^2=0.51\pm0.14(stat.)\pm0.07(syst.)$ which deviates from the perturbative QCD prediction of one by more than three standard deviations.  This significant disagreement suggests that either the decay is not dominated by single-virtual-photon annihilation as expected, or the QCD prediction does not apply to data in this energy region.  Because charged $\rho$'s are involved, $\rho^+\rho^-$ cannot be produced via two-virtual-photon annihilation unless there is significant final state interaction.  Given the possible relevance to potential similar effects in $B^0\to\rho^+\rho^-$ process, which is crucial for the determination of angle $\alpha$ through $CP$-violation in this mode, the understanding of the observed $e^+e^-\to\rho^+\rho^-$ decay amplitudes could have broader implications that should be kept in mind. 

\section{CONCLUSION} \label{sec:conclusion}
Over the past nine years of operation, the high luminosity data collected at the B factories have reopened several interesting areas for hadron physics.  Belle has provided a high statistics measurement of neutral $\pi$-pair production in two-photon physics.  BaBar has exploited initial-state radiation to study $e^+e^-$ annihilations at $E_{CM}$ from threshold to $\sim5$ GeV, to study the resonant substructure, and to contribute to the hadronic cross section measurements that are important in improving our theoretical understanding of $a_\mu$ and $\alpha(M_Z)$.  BaBar also has measured the cross sections and studied angular amplitudes of $p\bar{p}p\bar{p}$ and $\rho^+\rho^-$ in $e^+e^-$ annihilations.  The amplitude results in $e^+e^-\to\rho^+\rho^-$ are puzzling.  More interesting results from light hadron production are yet to come from the $B$ factories.

\end{document}